\documentclass{ws-procs9x6}

\def\GeV{\rm GeV}
\def\msbar{\overline{\rm MS}}

\begin{document}

\title{Parton Distributions -- DIS06.}

\author{\vspace{-0.9cm}R.S. Thorne\footnote{\vspace{-0.6cm}\uppercase{R}oyal 
\uppercase{S}ociety \uppercase{U}niversity \uppercase{R}esearch 
\uppercase{F}ellow.}}

\address{\vspace{-0.1cm}Department of Physics and Astronomy,
University College London\\
Gower St, 
London, WC1E 6BT, UK}

\maketitle

\abstracts{\vspace{-1.1cm}I discuss the current status of parton 
distributions. I outline 
the wide variety of different parton distributions available, and 
highlight which are either necessary or suitable for use at present.}

\vspace{-1.1cm}
There are a large number of different parton distributions. 
If we consider the different types of particle which are 
partons we start with  the quark model valence 
partons $u_V(x,Q^2)$ and $d_V(x,Q^2)$. However, these carry only 
$50\%$ of the proton's momentum: 
there are also $\bar u(x,Q^2)$ and $\bar d(x,Q^2)$, which are 
not the same\cite{E866};
gluons $g(x,Q^2)$ which carry over $30\%$ of the 
momentum; $s(x,Q^2)$ and $\bar s(x,Q^2)$, with the 
possibility\cite{NuTeVdimuon} that 
$s(x,Q^2) \not= s(x,Q^2)$;
$c(x,Q^2)$ and $b(x,Q^2)$, which are 
perturbatively generated 
(there could be intrinsic contributions\cite{intrinsic} at large $x$); 
and at some level  
isospin violation, i.e. $u^p(x,Q^2) \not= d^n(x,Q^2)$, 
$u^p(x,Q^2) \not= d^n(x,Q^2)$ -- automatically the case with 
QED corrections\cite{MRSTQED}, which also lead to 
a $\gamma(x,Q^2)$ distribution.  
Overall there are $6-16$ different parton distributions, but some 
are very small and are often not needed.
There is another way of counting, i.e. the different sets of parton 
distributions from different prescriptions: 
LO, NLO, or NNLO in $\alpha_S$; with resummation 
corrections or allowances for higher twist; using 
$\msbar$, DIS, or potentially other factorization schemes;   
fixed-flavour (FFNS), zero-mass variable-flavour
(ZM-VFNS) and general-mass variable-flavour number scheme
GM-VFNS, and even different versions of the last. This freedom, and 
choices in data sets used, cuts applied, ways 
to treat errors, {\it etc.} lead to a staggering array -- 
CTEQ4A1, CTEQ4A2\cite{CTEQ4} ... CTEQ5HJ\cite{CTEQ5}, ... CTEQ6\cite{CTEQ6}
..., MRST98\cite{MRST98} ... MRST03c\cite{MRSTerror2}, ... 
MRST04QED\cite{MRSTQED},
MRST04\cite{MRST04}..., Alekhin00\cite{Alekhin1}, Alekhin03\cite{Alekhin2}, 
GRV98\cite{GRV98}, Fermi02\cite{GKK}...,
ZEUS\cite{ZEUS}, ZEUS-ZJ\cite{ZEUSJ}, H1\cite{H1}, Botje\cite{Botje}
and many more. Are all of these different sets really necessary?  
This is a complicated and controversial question.

One restriction is very easy to impose -- 
many (still used) partons are simply out of date.
Unless there is a very good reason, one should not use pre-2000 
parton distributions.
The available data have improved a great deal since then, 
particularly the HERA structure 
functions\cite{H1,H1ZEUSdata} and Tevatron jets\cite{jets}.
Also, some older partons have minor bugs.

\begin{figure}[ht]
\vspace{0.3cm}
\centerline{\hspace{-0.5cm}\epsfxsize=1.8in\epsfbox{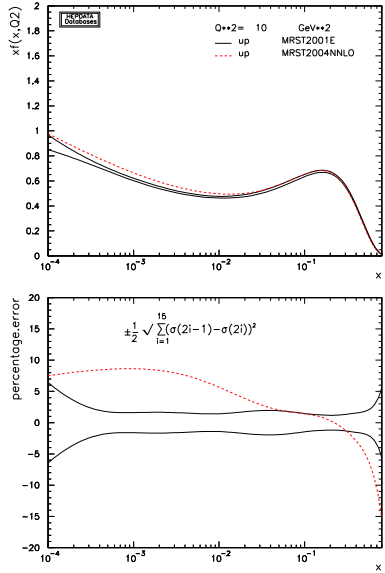}}   
\vspace{0.2cm}
\caption{Comparison of the NLO up distribution with the NNLO up distribution.
  \label{fig1}}
\vspace{-0.5cm}
\end{figure}

Consider the order of the partons. 
LO requires $\alpha_S(M_Z^2) \sim 0.130$, and 
the $\chi^2$ from a global fit is far inferior to that at NLO and 
NNLO\cite{MRSTNNLO}.
The LO partons are qualitatively different from 
NLO and NNLO partons in some 
regions, in particular the gluon is much bigger at 
small $x$ due to important 
corrections in the splitting functions. 
This can cause misleading conclusions on the evidence for 
saturation  {\it etc.}. 
Such results derived from LO partons should be treated with care.
The default has long been NLO, 
but the 
NNLO coefficient functions for structure functions have long been 
known\cite{CF} and 
the splitting functions are now complete\cite{NNLOs}. 
These improve the quality of the fit 
slightly\cite{MRSTNNLO} and reduce $\alpha_S$. 
A big change in the partons can occur when going from 
NLO $\to$ NNLO, as seen in Fig.~\ref{fig1}. 

To perform an absolutely correct NNLO fit we need 
both exact NNLO splitting functions and cross-sections. 
The NNLO 
Drell-Yan cross-sections have recently been calculated as a function of 
rapidity\cite{NNLODY}, leading to a decrease in the  
sea quarks. The one remaining gap is  
the NNLO corrections to jet production in 
$pp (\bar p)$ collisions. However, the NLO 
corrections themselves are not large --
at central rapidities they are $\leq 10\%$, similar to the size of 
the correlated errors. There are also some NNLO estimates, 
i.e. the leading threshold corrections, which are  
expected to be a significant 
component of the total\cite{kidonakis}
(there are issues concerning the application within a given jet definition).
These give a flat, small $3-4\%$ correction, which 
is consistent with what we already see 
at NLO and is  much smaller than the systematic errors on the data.
Hence, it seems perverse to leave the jet data out of a NNLO fit due to 
the lack of the full NNLO hard cross-section. 

For a  full NNLO fit we also require a rigorous treatment of heavy quark 
thresholds, which is now available\cite{nnlovfns}. 
Therefore, an essentially full NNLO determination of partons is 
possible. 
Of course, this is the best way to test our understanding of the partons in 
terms of QCD, but we only know a limited number of cross-sections at NNLO.
Processes with two strongly interacting particles are largely 
completed -- DIS coefficient functions,
$pp(\bar p) \to$ $\gamma^{\star}, W, Z$
(including the rapidity distribution), $H, A^0, WH, ZH$.
For many other final states the NNLO cross-sections are not known and 
NLO is still more appropriate. Moreover,   
resummations may be important even beyond NNLO in some 
regions, as may higher twist. 

There is the issue of factorization schemes. 
In practice cross-sections are calculated in  
$\msbar$ scheme, so we use $\msbar$ parton 
distributions. However, DIS-scheme can be more useful for relating 
partons to real physical results, 
e.g. it is easier to fit the Tevatron jet data in the DIS scheme\cite{KK}, 
or viewed differently the large high-$x$ gluon required at 
NLO and NNLO in $\msbar$ scheme can be determined from the scheme 
dependence\cite{MRST04}. Schemes other than $\msbar$ are 
valuable in this type of context, but at present rarely used.

There are also partons corresponding to different
prescriptions for heavy flavours. FFNS is intrinsically inferior to VFNS -- 
it does not sum $\ln Q^2/m_H^2$ terms in the perturbative expansion
and at high scales this can lead to inaccuracies. Moreover,
it is often necessary to have heavy flavour partons due to the lack of 
mass effects in the known cross-sections.
Nevertheless, FFNS partons are also sometimes needed because some hard 
cross-sections are only calculated in this scheme\cite{harris,mcatnlo,Herwig}. 
However, in this case the treatment must be correct, and is often not so. 
The NLO ($O(\alpha_S^2)$) coefficient functions 
for heavy flavour in DIS are calculated\cite{nlocalc} 
in a renormalization scheme where the 
coupling $\alpha_S$ is fixed at 3 flavours. The partons have to be defined 
in the same way, otherwise there is double counting of 
$\alpha_S^2\ln^2(Q^2/m_H^2)$, terms which can lead to large 
errors\cite{MRSTFFNS}. Also, there are no FFNS coefficient functions at NNLO.
This absence is particularly important since NNLO FFNS contains terms of the 
form $\alpha_S^3 \ln^2(Q^2/m_H^2)$.

\begin{figure}[ht]
\vspace{-0.65cm}
\centerline{\hspace{-0.6cm}\epsfxsize=2.55in\epsfbox{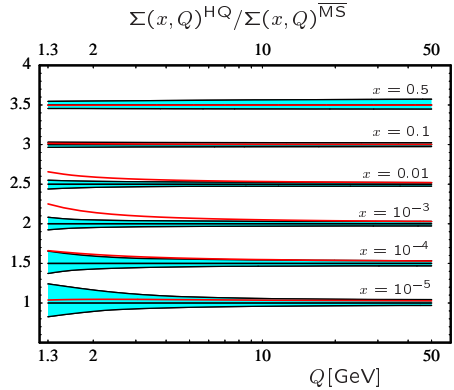}}   
\vspace{-0.3cm}
\caption{Comparison of the CTEQ6M singlet distribution with uncertainties 
and the CTEQ6HQ singlet distribution.
  \label{fig2}}
\vspace{-0.55cm}
\end{figure}

At the other extreme we have the ZM-VFNS. Here the 
terminology {\it scheme} is misleading. It usually means a 
different way of arranging the complete calculation. 
In this case there is an overall error of $O(m_H^2/Q^2)$. 
In my opinion ZM-VFNS is not  
useful. At high scales we are often in the limit where charm and bottom are 
effectively massless and a GM-VFNS is identical to the ZM-VFNS. However, 
the partons are obtained from fitting to data 
in regions where $O(m_H^2/Q^2)$ corrections are important, and
ignoring these leads to incorrect partons at all $Q^2$
in the ZM-VFNS. In Fig. \ref{fig2} we see the difference between the 
GM-VFNS CTEQ6HQ partons\cite{CTEQ6HQ}
and the ZM-VFNS CTEQ6 partons with their (conservative) uncertainties. 
At NNLO the partons become discontinuous at the transition 
points, indeed $c(x,Q^2)$ 
at $m_c^2$ is negative, and at this order we certainly 
need a GM-VFNS. 
If for some process GM-VFNS coefficient functions are not known, the 
error of $O(m_H^2/Q^2)$ from using the GM-VFNS partons is no worse than the 
permanent error from using ZM-VFNS. 
At worst we can input kinematic 
constraints into coefficient functions. 
There are a variety of definitions of a 
GM-VFNS\cite{vfns}, 
but they generally agree on fundamentals. Each choice is superior to 
ZM-VFNS and to FFNS.  
However, most are not defined in detail up to NNLO\cite{nnlovfns}, and
there are some lingering differences.

\begin{figure}[ht]
\vspace{-0.6cm}
\centerline{\hspace{-0.8cm}\epsfxsize=1.95in\epsfbox{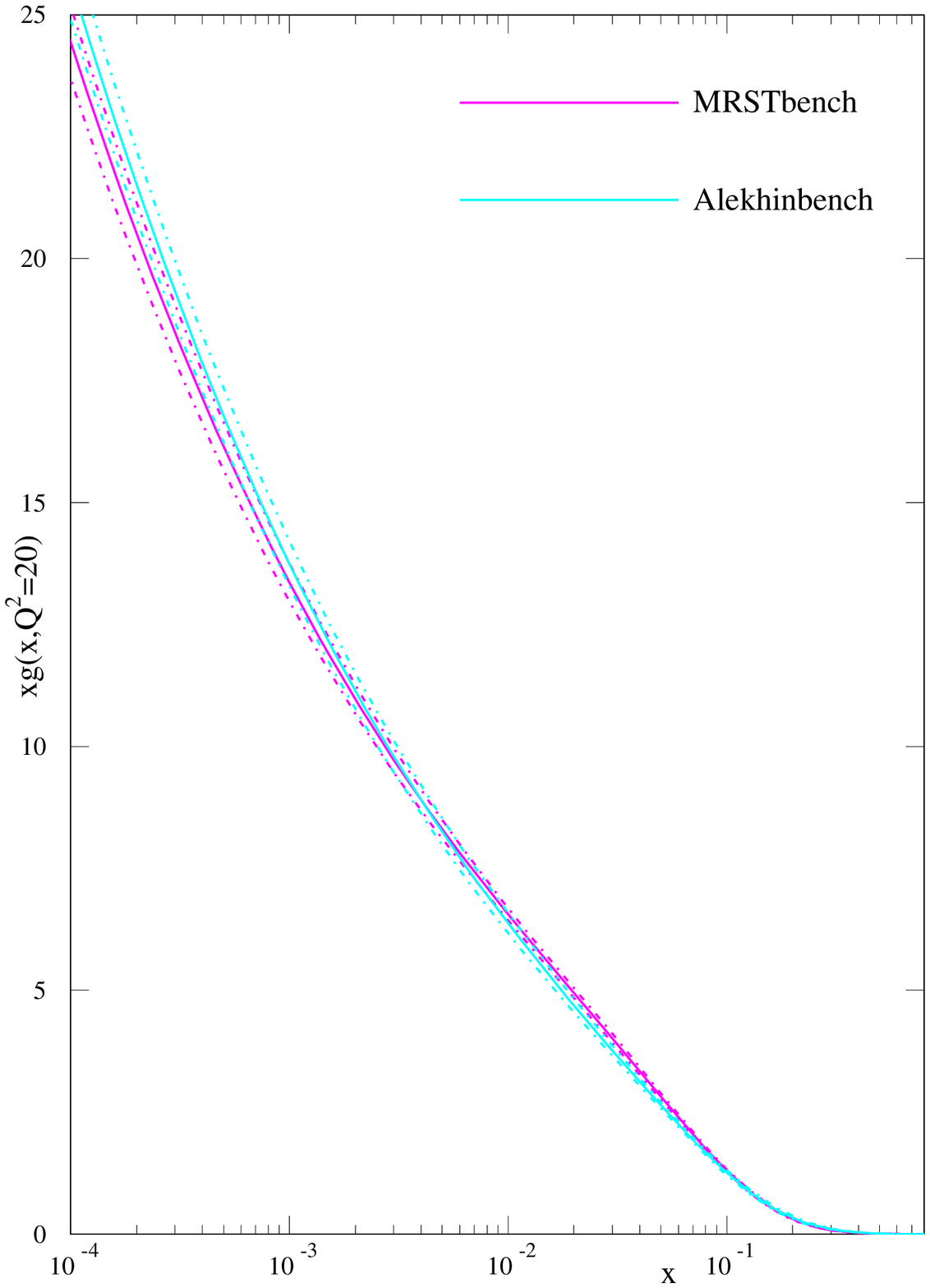}
\epsfxsize=1.95in\epsfbox{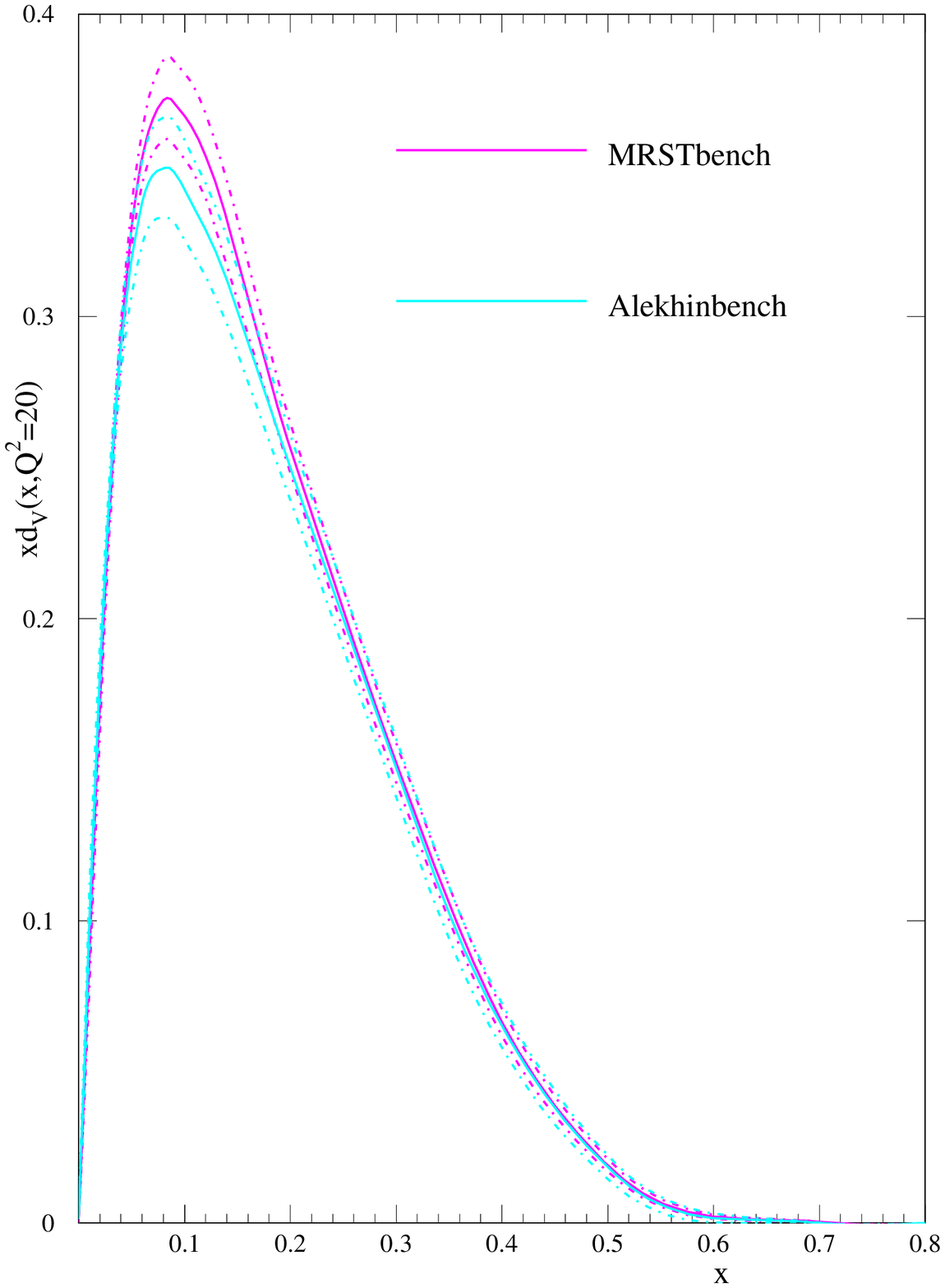}}   
\vspace{-0.2cm}
\caption{Comparison of the benchmark gluon distributions and  
$d_V$ distributions.
  \label{fig3}}
\vspace{-0.6cm}
\end{figure}

For a given theoretical prescription we still 
have a wide choice of partons. 
It is obvious that some competition is necessary, but
not all partons are equal  --
some are, in some sense, incorrect. There are a  
variety of reasons for this -- bugs in programs, incorrect theoretical 
approach (e.g. wrong coupling for flavour scheme), approximations to complete
theoretical approach,  or region of applicability, e.g. MRST03c partons are 
only suitable within the region of cuts on the data fit. 
The error is sometimes small, but can be the size of the intrinsic 
uncertainty or greater. If so, such partons should not be used. 
Indeed, NNLO is often still in the approximate stage.

There is also the issue of the treatment of experimental errors. 
As an exercise for the HERA-LHC\cite{HERALHC} 
workshop, partons were produced from fits 
to H1, ZEUS, NMC\cite{NMC} and BCDMS\cite{BCDMS} structure function data 
for $Q^2 > 9 \GeV^2$
using ZM-VFNS and a common form of parton inputs at 
$Q_0^2 =1 \GeV^2$ -- clearly very conservative. Partons were 
obtained using the rigorous treatment of all systematic errors 
(labelled Alekhin)
and using the simple quadratures approach (labelled MRST), both using 
$\Delta \chi^2=1$ to define the limits of uncertainty. 
As seen in  Fig.~\ref{fig3} there are 
small differences in the central values and similar errors, i.e.
the two sets are fairly consistent.
Even so, the full treatment of systematic errors is presumably better, but 
perhaps it is not so straightforward. 
Consider the averaged H1-ZEUS data sets\cite{glazov}, 
where the systematics of one data set 
can be significantly reduced by fitting to the other set.
The averaged data set is much more precise with very small
systematic errors. At the HERA-LHC workshop a 
comparison was made of a fit to both data
sets and a fit to the averaged data set\cite{Mandy}.
The partons resulting can differ by more than the uncertainty in each, 
and the movement of the data relative to the theory was different in each 
case. Data can move relative to theory due to
systematic uncertainties, but in reality this may be due to failures in 
theory rather than due to the central values for the data being 
incorrect. The conventional approach to systematic errors assumes we 
fit data to a perfect theoretical model with some unknown parameters, whereas 
in fact we are testing 
QCD at some order, and it may be slightly lacking. 
It is always best to remember this and try to minimize systematic errors. 
This makes the averaged H1/ZEUS data set very desirable. It is easier to 
understand and trust dominant statistical errors.

\begin{figure}[ht]
\vspace{-0.5cm}
\centerline{\hspace{-0.5cm}\epsfxsize=2in\epsfbox{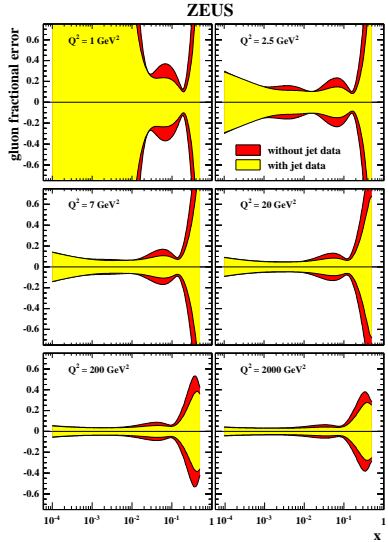}}   
\vspace{-0.3cm}
\caption{Comparison of the uncertainty on the ZEUS gluon distribution with 
and without the inclusion of their jet cross-section data.
  \label{fig4}}
\vspace{-0.5cm}
\end{figure}

\begin{figure}[ht]
\vspace{-0.7cm}
\centerline{\hspace{-0.8cm}\epsfxsize=1.9in\epsfbox{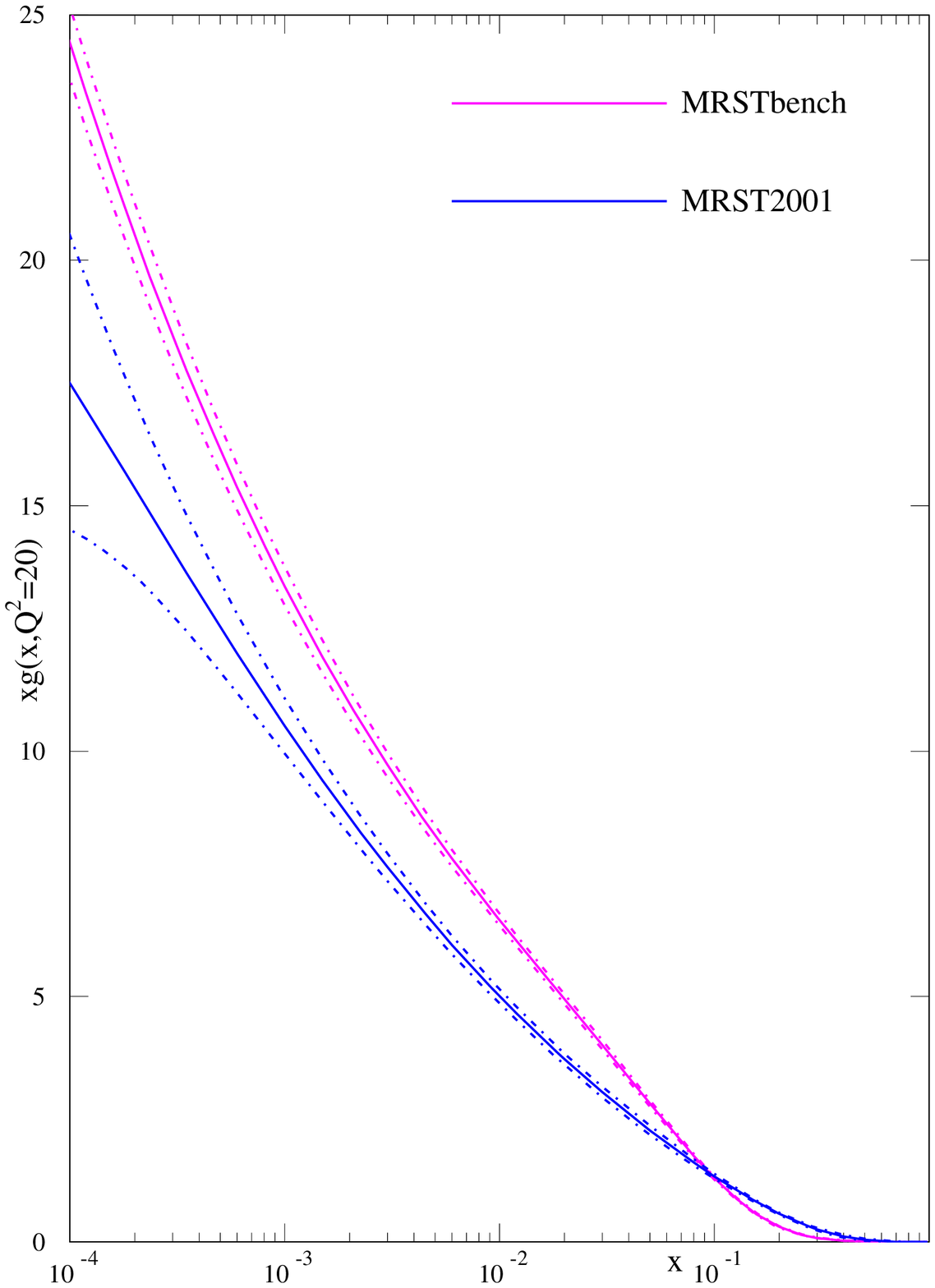}
\epsfxsize=1.9in\epsfbox{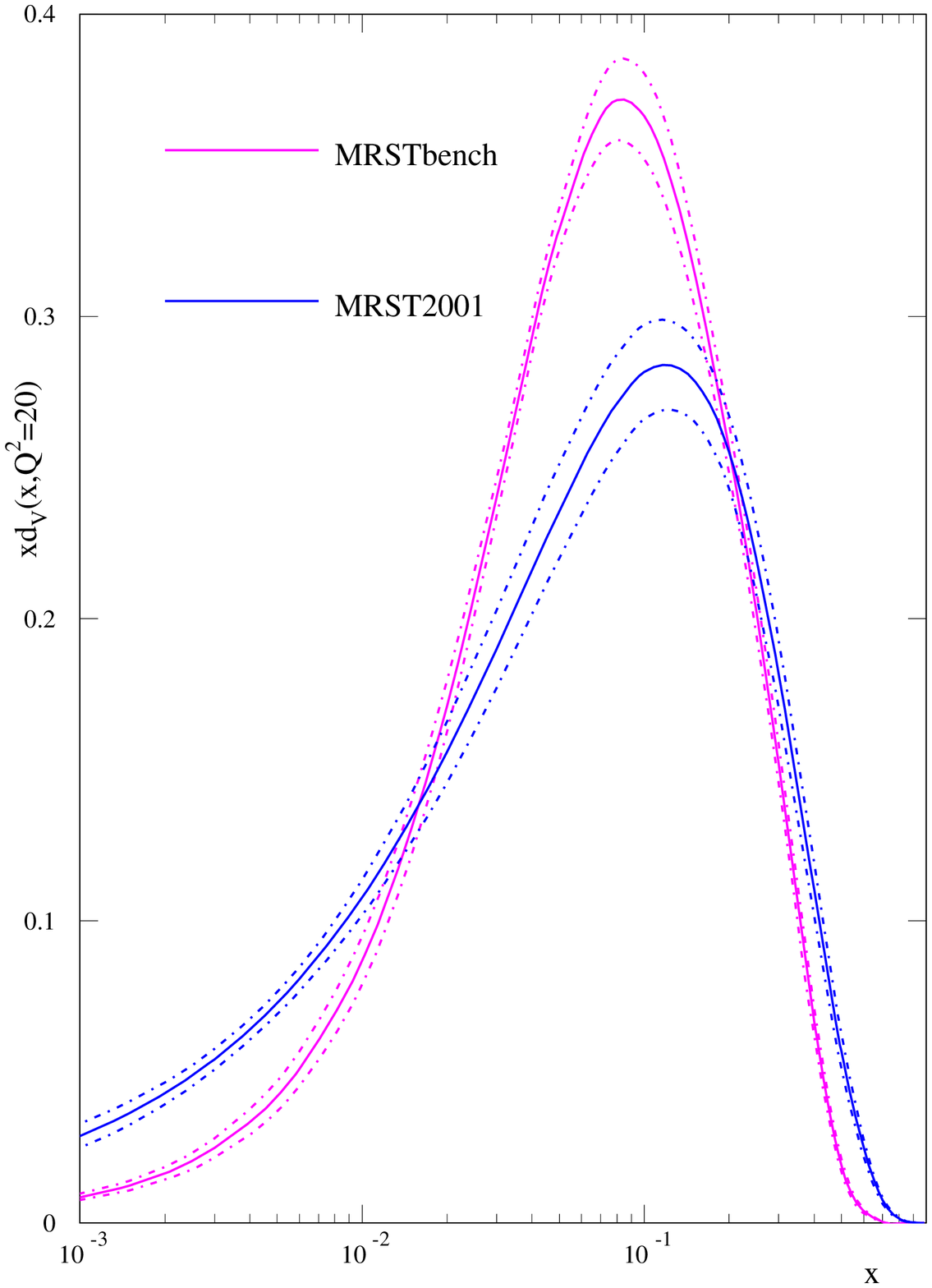}}   
\vspace{-0.2cm}
\caption{Comparison of the benchmark gluon distribution and  
$d_V$ distribution with the corresponding MRST2001E partons. 
  \label{fig5}}
\vspace{-0.4cm}
\end{figure}

We should also include as many data as possible in order 
to determine the partons, 
e.g. we see in Fig.~\ref{fig4} the reduced uncertainty in 
the ZEUS partons when 
including their own jet data\cite{ZEUSJ}.
The central values can also move, but do not do so much in this case. As a
more dramatic example we consider  the HERA-LHC benchmark partons and
investigate how these compare to partons obtained from a global fit
(the MRST01 partons\cite{MRSTerror1}), where the uncertainty is 
determined using $\Delta \chi^2=50$. There is an 
enormous difference in the central values,
sometimes many $\sigma$, as seen in Fig.~\ref{fig5}. 
The uncertainties are similar using 
$\Delta \chi^2=1$ compared to 
$\Delta \chi^2=50$ with approximately twice the data. 
Moreover, $\alpha_S(M_Z^2)\!=\!0.1110\pm 0.0015$
from the benchmark fit compared to $\alpha_S(M_Z^2)\!=\!0.119\pm 0.002$. 
Something is clearly seriously wrong in one of these analyses, and 
I am very confident that it is the benchmark fit. It fails when compared 
to most data sets not included, and not all can be 
unreliable. Partons should be constrained by all 
possible reliable data. 
The benchmark fit partons are extreme, but some other partons frequently used 
are similar in terms of the quantity of data fit, but many input 
implicit constraints from 
elsewhere. Also, for the global fit $\Delta \chi^2=1$ is not reliable. 
There must be something better than $\Delta \chi^2 = 50(100)$ or the offset 
method\cite{ZEUS}, but we are not yet sure what that is. 
The problems are partially due to the strict 
incompatibility of different data sets. Systematic errors are 
difficult to understand and not usually Gaussian in nature.
Also our theory is never perfect -- it is not
simply a matter of tying down unknown constants.
There are corrections possible at low $Q^2$, small $x$ and high
$x$. Indeed, comparing different sets of partons, one finds that the 
gluon is still very uncertain at low $x$ and $Q^2$, even though
all partons are fit to the same small-$x$ HERA data.
The additional constraint from a direct measurement of
$F_L(x,Q^2)$ would help this situation\cite{Ringberg}.

To conclude, there are many types of partons, and although 
some may be ignored, a variety is needed for the full range of 
applications and to estimate the uncertainties due to different assumptions 
(though one should be wary of partons that have only a limited set of 
constraints from data).
We need different prescriptions for heavy flavours (though not
ZM-VFNS), different factorization schemes and different orders. As a test of 
QCD, NNLO is preferable, and we are now obtaining 
reliable NNLO partons.  
We sometimes need partons for special occasions, e.g. to investigate 
the NuTeV $\sin^2 \theta_W$ anomaly\cite{NuTeVst}. There are many 
available, with QED corrections\cite{MRSTQED}, isospin 
violation\cite{MRSTerror2}, $s(x,Q^2) \not= \bar s(x,Q^2)$\cite{CTEQs} 
{\it etc.}. We also need to determine whether resummations at small or 
large $x$, higher twist or other theoretical corrections are 
important in some regions. There is much activity in these areas, 
and hopefully it will very soon provide concrete results.

\end{document}